\documentclass[11pt]{article}
\usepackage{moriond,epsfig}

\def\be{\begin{equation}}
\def\ee{\end{equation}}
\def\bea{\begin{eqnarray}}
\def\eea{\end{eqnarray}}

\begin{document}
\vspace*{4cm}
\title{PROSPECTS FOR THE SEARCH FOR HIGGS BOSONS WITH \break VECTOR BOSON FUSION PROCESSES AT THE LHC}

\author{I. ROTTL\"{A}NDER }

\address{Physikalisches Institut, University of Bonn, Nussallee 12, \\53115 Bonn, Germany}

\maketitle\abstracts{
The search for the Higgs boson is one of the main physics goals of the Large Hadron Collider (LHC) and its two multi-purpose experiments, ATLAS and CMS.
Vector boson fusion is the second largest production process for a standard model Higgs boson at the LHC and offers excellent means for background suppression. 
This paper gives an overview of the prospects of Higgs boson searches using vector boson fusion at the LHC. 
For a standard model Higgs boson, the decay channels ${\rm H\rightarrow \tau \tau}$, ${\rm H\rightarrow WW}$ and ${\rm H\rightarrow \gamma \gamma}$ are discussed. The discovery potential in the framework of the MSSM is summarized.} %

\section{The Large Hadron Collider}
The \begin{it} Large Hadron Collider \end{it}(LHC) is a proton-proton collider, which is currently being build at CERN, Geneva.
The first physics run at the design center-of-mass energy of 14 TeV is expected for 2008. At first, the LHC will operate at low luminosity ($2\cdot10^{33}$cm$^{-2}$s$^{-1}$) and will later increase its luminosity to the design value of $10^{34}$cm$^{-2}$s$^{-1}$. ATLAS \cite{ATDR} and CMS \cite{CTDR}, the two multi-purpose experiments at the LHC, are designed to investigate a wide range of physics.


\section{Vector boson fusion (qqH-production)}
In the vector boson fusion process, two weak bosons are radiated off the incoming quarks and merge to give the Higgs boson. Vector boson fusion is the second largest production process for a standard model Higgs boson at the LHC.\cite{Hahn}  It has a clear signature which can be used to efficiently suppress the background.
\subsection{Tagging Jets}
One of the most characteristic features of vector boson fusion processes are the \begin{it}tagging jets \end{it}that are produced from the quarks that are scattered off the merging heavy vector bosons. These jets typically have high ${\rm p_T}$ and lie in different hemispheres in the forward- and backward region of the detector.\cite{Zeps}
Very forward jets may not be fully contained in the calorimeter, but partially lie in the beam pipe.
Moreover, underlying event \footnote{Additional interactions from the  same proton-proton collision as the hard event.} and  pile-up \footnote{Events from other proton-proton collisions than the hard collision, which might also arise from a later or earlier bunch crossing. For the low luminosity phase, two to three collisions per bunch crossing are expected. } lead to additional energy depositions in the calorimeters.
A reliable performance and good understanding of the forward calorimeters as well as efficient energy clustering and jet finding algorithms are therefore needed [Fig. 1].

\begin{figure}[tp]
\begin{minipage}[b]{.43\linewidth} 
\centerline{\epsfig{file=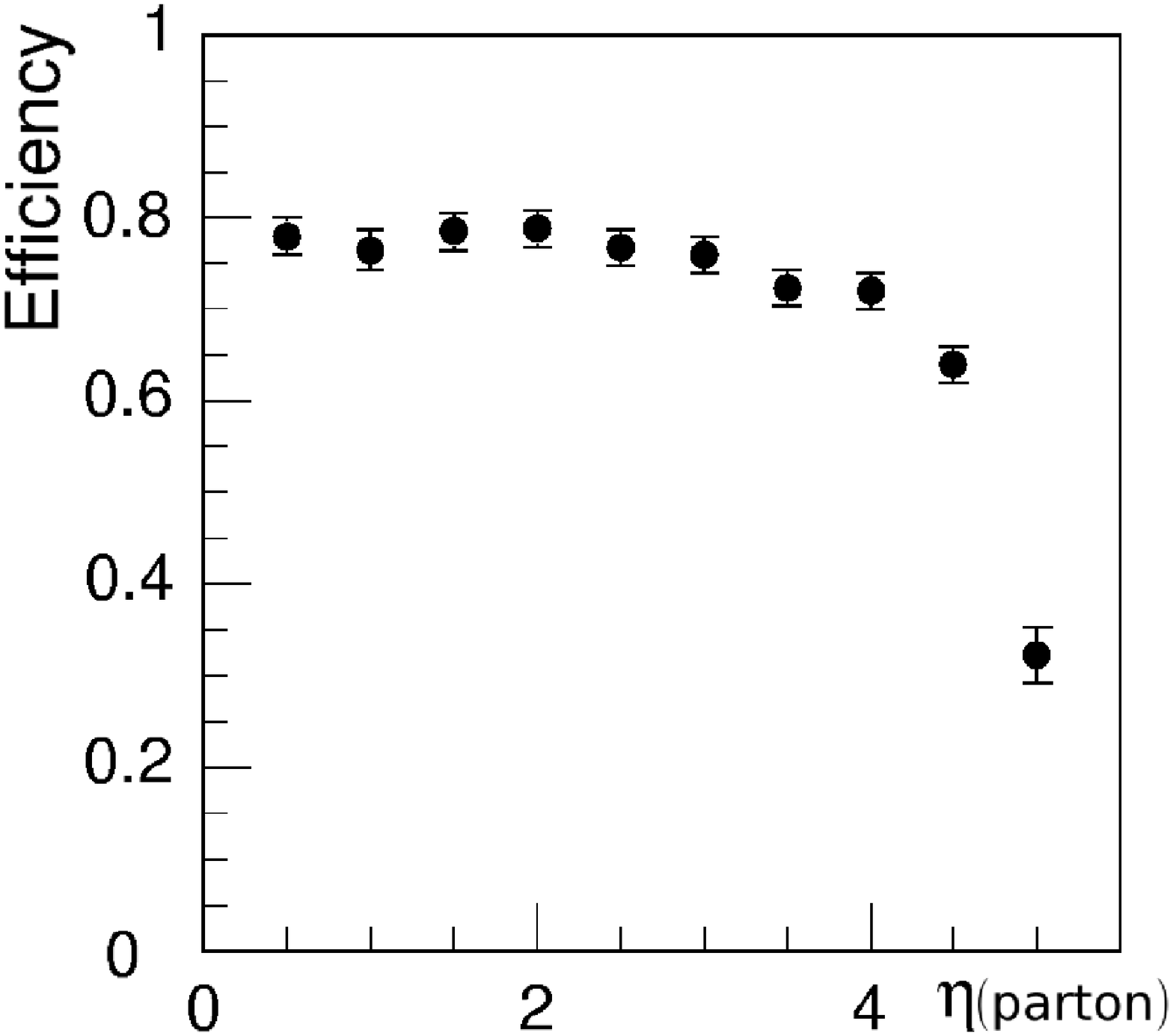, width=4.98cm}}
\caption{Efficiency for reconstructing a tag jet with ${\rm p_T}$$>$20GeV from a parton with ${\rm p_T}$$>$20GeV at ATLAS.$^6$ Calorimeter coverage ends at $\eta= 5$.}
\end{minipage}
\hspace{.05\linewidth}
\begin{minipage}[b]{.47\linewidth} 
\centerline{\epsfig{file=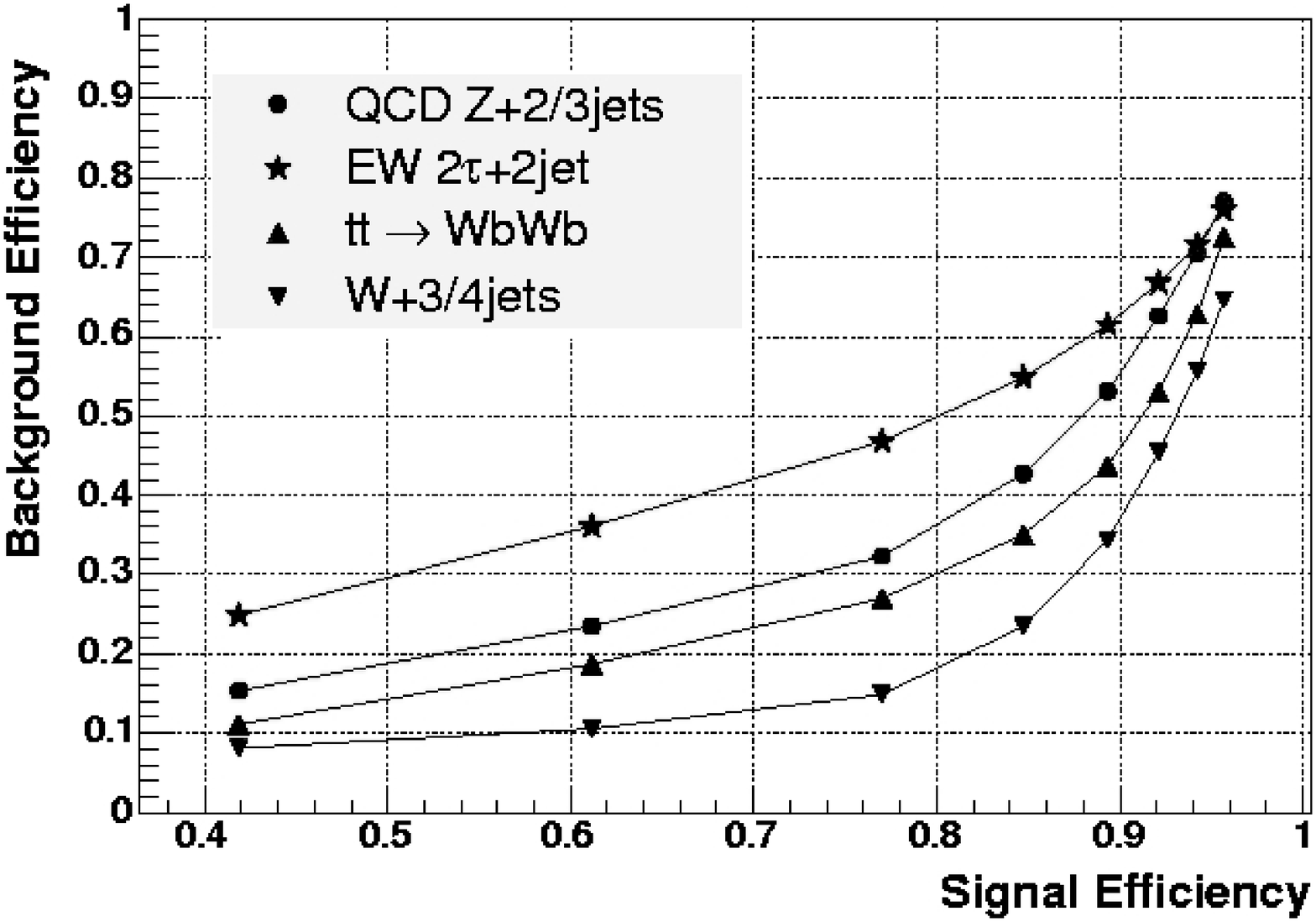, width=5.7cm}}
\caption{ Central jet veto efficiencies for four background processes and the ${\rm H\rightarrow \tau \tau}$ -signal for eight ${\rm p_T}$-cuts on the central jet (${\rm p_T}$$<$10,15,...45GeV, cor\-responding to the eight signal efficiencies) at CMS.$^{9}$}
\end{minipage}
\end{figure}

\subsection{Central Jet Veto}
The decay products of the Higgs boson typically lie in the central detector region.\cite{Zeps} 
Since there is no colour flow between the quarks in the vector boson fusion process, jet production in the central region is suppressed. In contrast, central emission is favoured in QCD interactions which constitute important background processes at the LHC.\cite{Zeps,MJV}
A veto on additional jets in the central region is therefore a powerful discriminant between vector boson fusion and QCD background processes such as $\rm {t\bar{t}}$-production [Fig. 2].\cite{CTDR,AVBF}\\
Underlying event and pile-up may produce jets in the central region that do not originate from the vector boson fusion hard interaction. This may considerably alter the performance of the jet veto.
A good understanding of underlying event and pile-up levels 
is crucial for this reason.\\
Because of the large uncertainties of forward jet tagging and the central jet veto in the presence of pile-up, all analyses presented here assume the low luminosity case.


\begin{figure}[bp]
\begin{minipage}[b]{.45\linewidth} 
\centerline{\epsfig{file=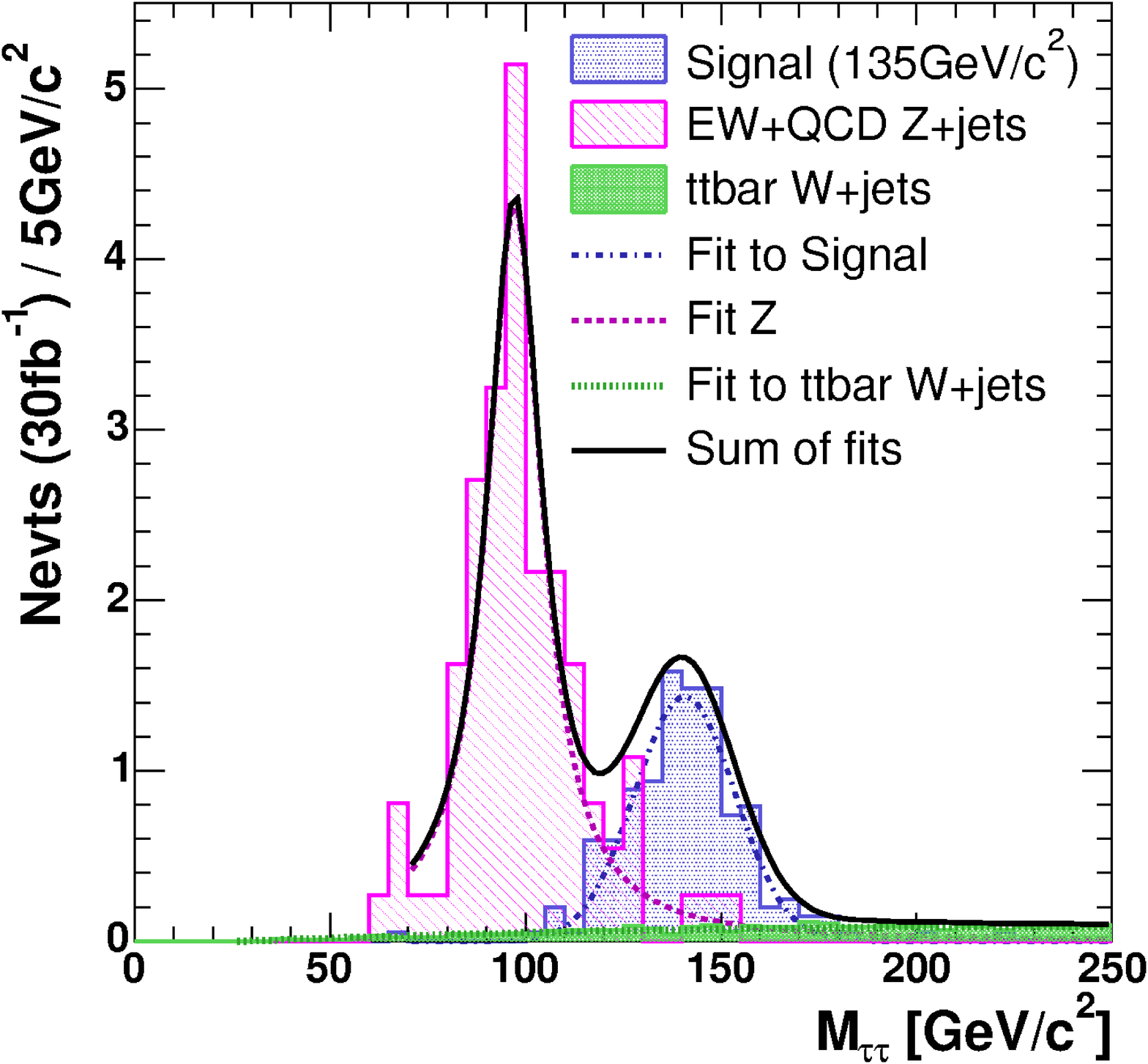, width=4.98cm}}
\caption{Reconstructed Higgs boson mass for ${\rm H}$$\rightarrow$$\tau$$\tau$$\rightarrow$lh after analysis cuts at CMS (30fb$^{-1}$).$^{9}$}
\end{minipage}
\hspace{.05\linewidth}
\begin{minipage}[b]{.45\linewidth} 
\centerline{\epsfig{file=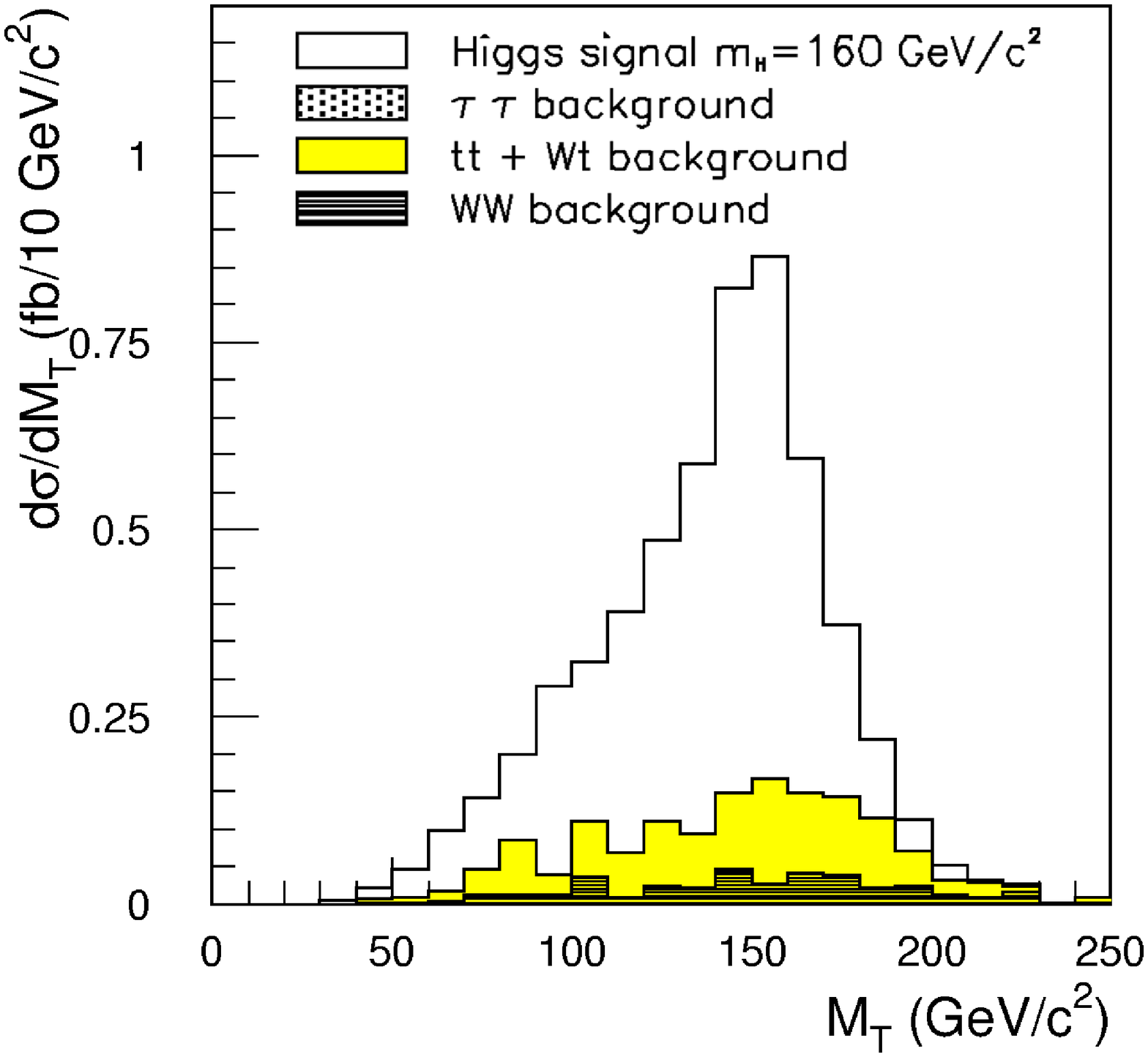, width=5.07cm}}
\caption{ Transverse mass distribution for ${\rm H}$$\rightarrow$${\rm W}$${\rm W}$$\rightarrow$${\rm e}$$\mu$ after analysis cuts at ATLAS.$^6$}
\end{minipage}
\end{figure}

\section{Search in the ${\rm H \rightarrow \tau\tau}$ decay mode}
The ${\rm H \rightarrow \tau \tau}$ channel \cite{AVBF,Ctau} allows access to a Higgs boson-fermion coupling in the decay. At the LHC, this is in the standard model otherwise only possible in the ${\rm H \rightarrow b\bar{b}}$ mode, which is experimentally challenging.\cite{CTDR,Jochen} 
For masses of the Higgs boson below ${\rm m_H= 140}$ GeV, its branching ratio to a $\tau$-lepton pair is around 4-8\% but drops at higher masses when the decay to heavy vector bosons opens up.\cite{hdecay}
For triggering, at least one high-${\rm p_T}$ lepton in the final state is needed. For this reason, the decay chains ${\rm H \rightarrow \tau\tau \rightarrow lh + 3\nu}$ and ${\rm H \rightarrow \tau\tau \rightarrow ll + 4\nu}$ are considered.
The main background process is irreducible ${\rm Z\rightarrow\tau\tau}$ -production from strong and electroweak processes. Also ${\rm t\bar{t}}$-production and WW-production in association with jets contribute.\\
In spite of the neutrinos in the final state, the reconstruction of the Higgs boson mass is possible by use of the collinear approximation.\cite{coll} Since the $\tau$-leptons obtain a large Lorentz boost due to the large mass and ${\rm p_T}$ of the Higgs boson, their decay products are emitted roughly in the direction of the original $\tau$. 
Exploiting momentum conservation in the transverse plane yields the 4-momentum vectors of the two $\tau$-leptons and thus the Higgs boson invariant mass [Fig. 3].\\
This channel is promising for a 5$\sigma$-discovery in a mass range of about 120 to 135 GeV with 30fb$^{-1}$. \cite{AVBF,Ctau} 
The Higgs boson will not be observable in this decay mode in inclusive searches.

\section{Search in the ${\rm H \rightarrow WW}$ decay mode}
The ${\rm H \rightarrow WW}$ channel \cite{AVBF,CW} gives clean access to the Higgs boson coupling to W-bosons both in production and in decay. The branching ratio rapidly increases with the Higgs boson mass and reaches values over 96\% at masses around ${\rm m_H\approx170}$ GeV. It drops to about 70\% when the decay to the heavier Z-boson opens up.\cite{hdecay}
\\
The semileptonic ${\rm H \rightarrow WW \rightarrow l\nu jj}$ and the purely leptonic ${\rm H \rightarrow WW \rightarrow l\nu l \nu}$ decay modes provide the required lepton for triggering. Especially for the semileptonic mode, large background contributions have to be suppressed. These include ${\rm t\bar{t}}$, W, Z, WW and ZZ-production in association with jets as well as QCD multijet production.\\
Since Higgs boson masses near the production threshold of the W-boson pair are considered, the decay products of the Higgs boson have relatively low ${\rm p_T}$. The collinear approximation is therefore not usable. Instead, the transverse mass ${\rm M_T=\sqrt{2\:p_{T,ll}\:p_{Tmiss}(1-\cos{\Delta \phi})}}$ is calculated [Fig. 4].\cite{AVBF} 
 In case of the ${\rm H \rightarrow WW \rightarrow l\nu jj}$ -decay, the Higgs boson mass can also be reconstructed 
by using the known W-boson mass as an input.\cite{CW}\\
With a cut analysis method, this channel is promising for a 5$\sigma$-discovery in a mass range of 125 to 190 GeV with 30fb$^{-1}$. Only 5fb$^{-1}$ are needed in the mass region of 150 to 190 GeV.\cite{AVBF}

\section{Search in the ${\rm H \rightarrow \gamma \gamma}$ decay mode}
The branching ratio of the standard model Higgs boson to two photons is at most 0.22\% at ${\rm m_H\approx125}$ GeV.\cite{hdecay} 
However, this decay channel has an excellent Higgs boson mass resolution.
The main background is photon production in association with jets. Processes with final state electrons and jets contribute if these particles are misidentified as isolated photons.\\ 
To achieve a final mass resolution of $\sigma_{\gamma \gamma}$=0.7\% \cite{Cgamma}, it is necessary to reconstruct the Higgs boson vertex to reduce
the photon direction uncertainty.
An efficient method for vertex reconstruction is to determine the longitudinal position of the highest ${\rm p_T}$ track in the event.\cite{Cgamma}\\
Using a cut analysis method, a significance of 2.5$\sigma$ can be achieved for a Higgs boson mass of 120 GeV with 30fb$^{-1}$. \cite{Cgamma} As a discovery channel, vector boson fusion ${\rm H \rightarrow \gamma \gamma}$ is not competitive with the inclusive mode which reaches significances above 5$\sigma$.\cite{ATDR,CTDR} 

\begin{figure}[tp]
\begin{minipage}[b]{.46\linewidth} 
\centerline{\epsfig{file=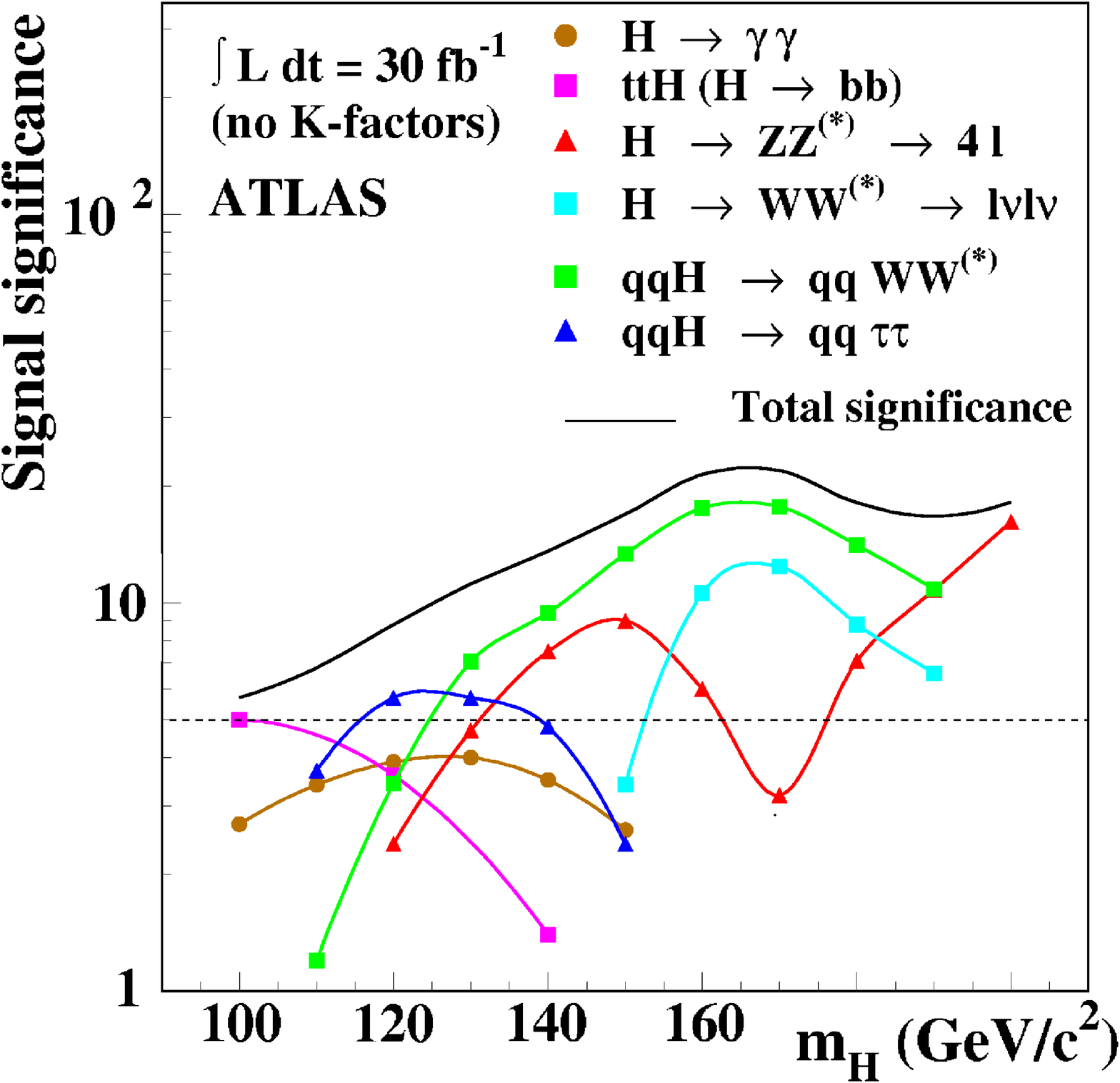, width=6.09cm}}
\caption{Discovery potential for a standard model Higgs boson in different search channels at ATLAS.$^{1}$}
\end{minipage}
\hspace{.05\linewidth}
\begin{minipage}[b]{.44\linewidth} 
\centerline{\epsfig{file=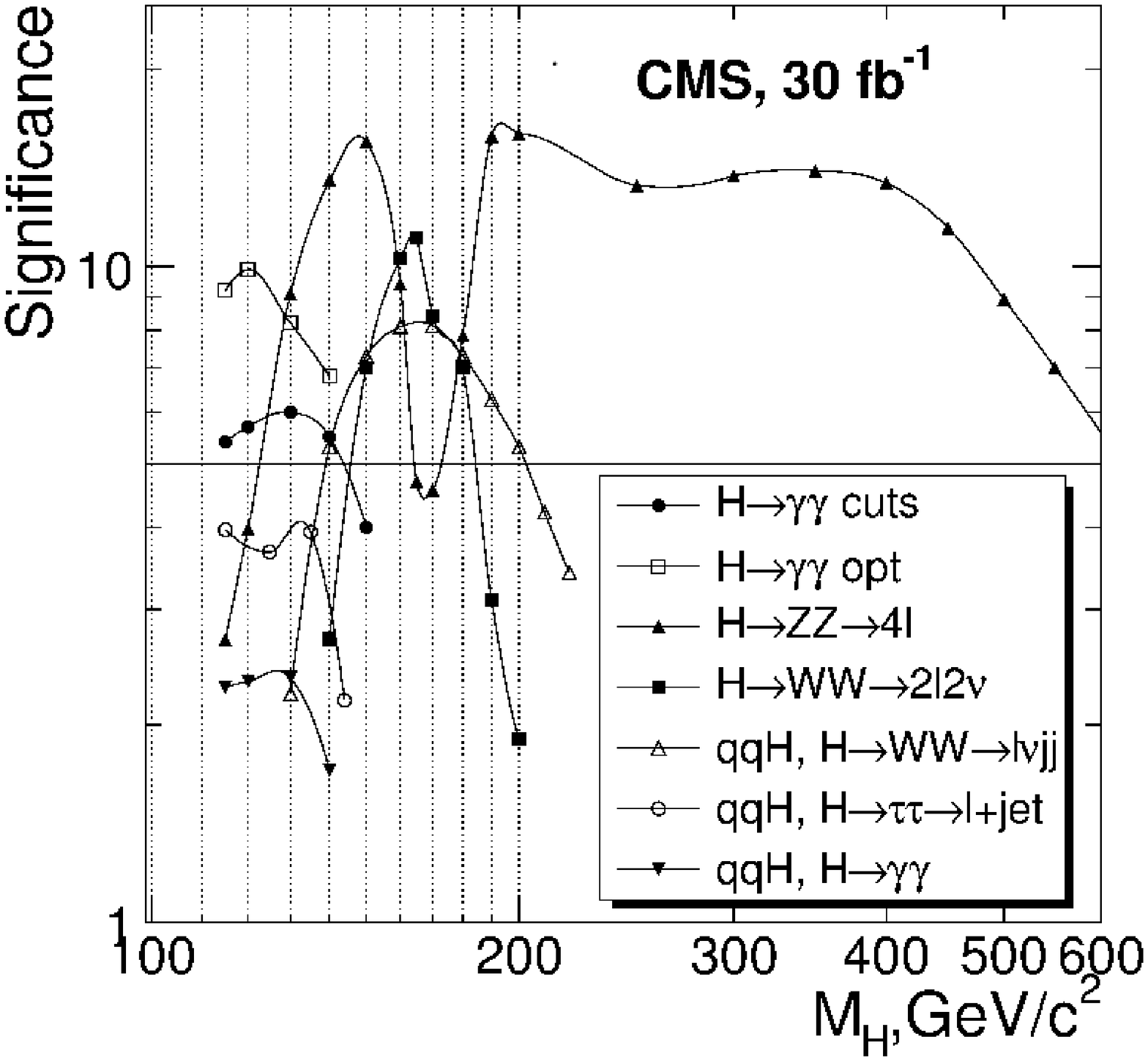, width=6.24cm}}
\caption{Discovery potential for a standard model Higgs boson in different search channels at CMS.$^2$}
\end{minipage}
\end{figure}

\section{Higgs bosons of the Minimal Supersymmetric Standard Model (MSSM)}
The MSSM features two SU(2) Higgs doublets with a total of five observable states, two neutral CP-even, one neutral CP-odd and one charged Higgs boson. 
The Higgs sector can  be described at tree level by two parameters, which are usually chosen to be the mass of the CP-odd Higgs boson, ${\rm M_A}$, and the ratio of the vacuum expectation values of the two Higgs doublets, ${\rm \tan \beta = \frac{v_1}{v_2}}$.\\
Only the CP-even Higgs bosons can be produced in vector boson fusion. At 30fb$^{-1}$, at least one of these Higgs bosons will be observable over the entire ${\rm M_A}$-$\tan \beta$ plane in the m$_{\rm H}^{\rm max}$, no-mixing, small $\alpha_{\rm eff}$ and the gluophobic benchmark scenario.\cite{MSSM,bench} 
Inclusive searches need higher integrated luminosities for complete coverage of the ${\rm M_A}$-$\tan \beta$ plane.\cite{MSSM}

\section{Conclusions}
Vector boson fusion offers an outstanding discovery potential for the standard model Higgs boson, especially in the low mass region between the LEP limit of 114.4 GeV \cite{LepLim} and 190 GeV [Fig. 5, 6].
In the supersymmetric framework of the MSSM, vector boson fusion processes have an excellent discovery potential over the whole parameter plane.

\section*{Acknowledgments}
This work was supported by the European Union/Marie Curie Actions.
Moreover, I would like to thank Alexandre Nikitenko and Markus Schumacher for helpful discussions.


\section*{References}
\begin{thebibliography}{99}

\bibitem{ATDR} ATLAS Collaboration, CERN-LHCC-99-014/CERN-LHCC-99-015.
\bibitem{CTDR} CMS Collaboration, CERN-LHCC-2006-001/CERN-LHCC-2006-021.
\bibitem{Hahn} T.~Hahn {\it et al}, {arXiv:hep-ph/0607308}.
\bibitem{Zeps} D.~L.~Rainwater and D.~Zeppenfeld,
  JHEP {\bf 9712} (1997) 005.

\bibitem{MJV}
 V.~D.~Barger, R.~J.~N.~Phillips and D.~Zeppenfeld,
  Phys.\ Lett.\  B {\bf 346} (1995) 106.
\bibitem{AVBF} 
S.~Asai {\it et al.},
  Eur.\ Phys.\ J.\  C {\bf 32S2} (2004) 19.

\bibitem{Jochen} J.~Cammin and M.~Schumacher, {ATLAS Internal Note} {\bf 2003/024}.

\bibitem{hdecay} 
A.~Djouadi, J.~Kalinowski and M.~Spira,
  Comput.\ Phys.\ Commun.\  {\bf 108} (1998) 56.

\bibitem{Ctau} C.~Foudas, A.~Nikitenko, M.~Takahashi, {CMS Note} {\bf 2006/088}.

\bibitem{coll} 
 R.~K.~Ellis, I.~Hinchliffe, M.~Soldate and J.~J.~van der Bij,
  Nucl.\ Phys.\  B {\bf 297} (1988) 221.
D.~L.~Rainwater, 
{arXiv:hep-ph/9908378}.
\bibitem{CW} H.~Pi {\it et al}, { CMS Note} {\bf 2006/092}.

\bibitem{Cgamma} M.~Dubinin {\it et al}, {CMS Note} {\bf 2006/097}.

\bibitem{MSSM} M.~Schumacher, {arXiv:hep-ph/0410112} and private communication.

\bibitem{bench} 
M.~Carena {\it et al},
  Eur.\ Phys.\ J.\  C {\bf 26} (2003) 601.

\bibitem{LepLim}
  R.~Barate {\it et al.}  
  Phys.\ Lett.\  B {\bf 565} (2003) 61.

\end{thebibliography}
\end{document}